# Structural, Surface Morphology and Magneto-Transport properties of Self Flux Grown Eu Doped $Bi_2Se_3$ Single Crystal

Rabia Sultana[1,2], Ganesh Gurjar[3], Bhasker Gahtori[1,2], Satyabrata Patnaik[3] and V.P.S. Awana[1,2*]

[1] National Physical Laboratory (CSIR), Dr. K. S. Krishnan Road, New Delhi-110012, India

[2] Academy of Scientific and Innovative Research (AcSIR), Ghaziabad-201002, India

[3] School of Physical Sciences, Jawaharlal Nehru University, New Delhi-110067, India

**Abstract**

Here, we report the effect of europium (Eu) doping in $Bi_2Se_3$ topological insulator (TI) by using different characterization techniques viz. X-ray diffraction (XRD), scanning electron microscopy (SEM) coupled with energy dispersive X-ray analysis (EDXA) and magneto-transport measurements. Good quality Eu doped $Bi_2Se_3$ ($Eu_{0.1}Bi_{1.9}Se_3$) single crystal is grown by the self flux method through the solid state reaction route. Single crystal XRD pattern displayed the high crystalline quality of the $Eu_{0.1}Bi_{1.9}Se_3$ sample along (00l) alignment whereas; the powder XRD confirmed the rhombohedral crystal structure without any impurity phases. SEM images exhibited a layered slab like structure stacked one over the other whereas; EDXA measurements confirmed the chemical composition of $Eu_{0.1}Bi_{1.9}Se_3$ sample. Further, the EDXA mapping showed the homogeneous distribution of Bi, Se and Eu elements. Temperature dependent electrical resistivity curves revealed a metallic behaviour both in the presence and absence of applied magnetic field. Magneto-transport measurements showed a decrease in the magneto-resistance (MR) value of the $Eu_{0.1}Bi_{1.9}Se_3$ sample (~32% at 5K) in comparison to the pure $Bi_2Se_3$ sample (~80% at 5K). For, $Eu_{0.1}Bi_{1.9}Se_3$ sample, a complex crossover between WL and WAL phenomenon was observed at lower applied magnetic fields, whereas the same was absent in case of the pristine one. Further, HLN (Hikami Larkin Nagaoka) fitted magneto-conductivity (MC) analysis revealed a competing weak anti localization (WAL) and weak localization (WL) behaviour. Summarily, in the present work we study the structural, surface morphology and magneto-transport properties of as grown $Eu_{0.1}Bi_{1.9}Se_3$ single crystals.

*Corresponding Author

Dr. V. P. S. Awana:  E-mail: awana@nplindia.org
Ph. +91-11-45609357, Fax-+91-11-45609310
Homepage: awanavps.webs.com

**Introduction**

The discovery of topological insulators (TIs), a new phase of quantum matter have surprised and fascinated physicists for about a decade [1-5]. In fact, it's extraordinary electronic properties along with a wide range of potential applications including the development of futuristic quantum computers is now among the hottest topics in physics research. In particular, TIs are characterized by a gapped bulk state and gapless surface/edge states which are further protected by time reversal symmetry (TRS) and have spin momentum locking property i.e., the conducting surface states of TIs are robust against time reversal invariant perturbations [1-11]. The existence of a topologically protected gapless surface state along with a single Dirac cone at the Γ point of the Brillouin zone is the most prominent property of a TI. Recent theoretical as well as experimental studies suggest that the breaking of TRS in TIs by magnetic doping opens up a gap in the spectrum of the surface states and hence generate massive surface carriers [12-14]. As reported, the bulk of a magnetically doped TI exhibits a long-range magnetic order both in the metallic and insulating phases through Van Vleck mechanism. Conversely, on the surface, such a long-range magnetic order can also be formed independent of the bulk magnetic ordering via the Ruderman-Kittel-Kasuya-Yosida (RKKY) exchange mechanism [12-16, 18, 25]. Consequently, doping three dimensional (3D) TIs ($Bi_2Te_3$, $Bi_2Se_3$, $Sb_2Te_3$) with transition metal elements (Cr, Fe, Mn, V, etc.) leads to the breaking of TRS and can bring about a long range ferromagnetic order either through Van Vleck / RKKY mechanism. Moreover, it has been shown that magnetic doping results into the realization of a variety of exotic topological properties such as the topological magneto-electric effect, quantum anomalous Hall (QAH) effect, imaging magnetic monopoles, and the Faraday and Kerr effects [12-25]. Accordingly, the magnetically doped TIs have received considerable attention and acts as one of the most exciting research area due to their striking topological phenomena (as mentioned above) arising from the breaking of TRS. Furthermore, an intrinsic TI shows the signatures of weak anti-localization (WAL) effect which is observed in the absence of any magnetic scattering. On the other hand, the electronic transport response of a magnetically doped TI exhibits a weak localization (WL) effect due to the surface gap opening induced by TRS breaking. There already exist some published reports on magnetically (Mn, Ni, Fe, Cr, V and Co) doped TIs, which focus on describing the effects of magnetic impurities, or discussing typical ferromagnetism on Dirac like conducting surface states in various TIs viz. $Bi_2Se_3$, $Bi_2Te_3$ and $Sb_2Te_3$ [25-44]. Till date, a very few reports have discussed on the magneto transport behaviour of Eu doped $Bi_2Se_3$ thin films [45, 46]. To the best of our knowledge, there has been no reports on the detailed physical property characterization of bulk Eu doped $Bi_2Se_3$ single crystals. Also, crystal growth itself is a very challenging task, as it depends upon the physical properties viz., melting point, volatile nature, solubility in water etc., of the materials under consideration. Furthermore, obtaining a high quality, reproducible single crystal along with interesting properties is the need of the hour. Keeping in view the importance of magnetically doped TIs, we report the structural, surface morphology and magneto-transport properties of Eu doped $Bi_2Se_3$ ($Eu_{0.1}Bi_{1.9}Se_3$) single crystal grown by the facile self flux method. As mentioned above, a magnetic dopant should destroy the topological surface robustness owing to the breaking of TRS, so we compare the results obtained for Eu doped $Bi_2Se_3$ ($Eu_{0.1}Bi_{1.9}Se_3$) single crystal with the pure $Bi_2Se_3$ single crystal results.

**Experimental Details**

Single crystals of $Eu_{0.1}Bi_{1.9}Se_3$ were grown using the standard flux free (self flux) method via the solid state reaction route. High purity (99.99%, Alfa Aesar) bismuth (Bi), selenium (Se) and europium (Eu) were taken as the starting materials. Stoichiometric mixture (~1gram) of the starting materials were taken and sealed in an evacuated quartz tube ($10^{-3}$Torr) as mentioned in our previously reported literature [44, 47]. Briefly, the sealed quartz tube containing the rectangular pellet was kept inside an automated tube furnace and heated to 950°C for 7.5hours (120°C/hour). The ampoule was kept at the same temperature (950°C) for 24 hours and then slowly cooled (2°C/hour) to 650°C. Further, a hold time of 48 hours was maintained at 650°C, followed by switching off the furnace to cool down naturally to room temperature. The detailed heat treatment diagram is displayed in Figure 1. The as grown $Eu_{0.1}Bi_{1.9}Se_3$ sample (~1cm) was taken out by breaking the quartz tube and mechanically cleaved for further characterizations.

The phase identification and crystalline nature of as grown $Eu_{0.1}Bi_{1.9}Se_3$ sample was carried out by employing Rigaku Miniflex II, Desktop X-ray Diffractometer (XRD) with Cu-K$\alpha$ radiation ($\lambda$=1.5418 Å). Scanning electron microscopy (SEM) coupled with energy dispersive X-ray analysis (EDXA) measurements were carried out using ZEISS-EVO MA-10 scanning electron microscope. The magneto - transport properties were measured using the Cryogenic System with fields up to 5Tesla and temperature down to 2K. The magneto-conductivity of $Bi_2Se_3$ and $Eu_{0.1}Bi_{1.9}Se_3$ TIs is understood on the basis of HLN (Hikami-Larkin-Nagaoka) equation.

**Results and Discussion**

The most straight forward and primary characterization technique for structural determination is the XRD. In order to identify the crystallinity, phase purity, crystal structure and lattice parameters, room temperature single crystal and powder XRD pattern were obtained for as grown $Eu_{0.1}Bi_{1.9}Se_3$ sample as shown in figure 2(a, b). Both (single crystal and powder) the XRD patterns were recorded using the Rigaku MiniFlex-II, desktop XRD with Cu - $K_\alpha$ radiation ($\lambda$=1.5418Å). The single crystal XRD pattern was taken on the silvery surface of mechanically cleaved $Eu_{0.1}Bi_{1.9}Se_3$ crystal (~4mm), whereas for powder XRD patterns several small crystals were thoroughly ground into powder form with the help of agate mortar and pestle.

Figure 2(a) depicts the single crystal XRD pattern obtained for Eu doped $Bi_2Se_3$ ($Eu_{0.1}Bi_{1.9}Se_3$) samples in the angular range of $2\theta_{min} = 10°$ and $2\theta_{max} = 80°$. The on surface XRD patterns exhibited well defined as well as sharp diffraction peaks, indicating good crystalline nature of the as synthesized $Eu_{0.1}Bi_{1.9}Se_3$ sample (Figure 2a). Apparently, the XRD pattern clearly shows that the synthesized $Eu_{0.1}Bi_{1.9}Se_3$ sample is single crystalline in nature and oriented in c-axis, similar to the pure $Bi_2Se_3$ sample.

Figure 2(b) shows the powder XRD pattern obtained for as grown Eu doped $Bi_2Se_3$ ($Eu_{0.1}Bi_{1.9}Se_3$) sample in the angular range of $2\theta_{min} = 10°$ and $2\theta_{max} = 80°$. Rietveld refinement of the obtained raw powder XRD (PXRD) data was performed with the help of FullProf Suite Toolbar software. The Rietveld refinement of the PXRD data confirmed that the Eu doped $Bi_2Se_3$ ($Eu_{0.1}Bi_{1.9}Se_3$) sample crystallized in rhombohedral crystal structure with $R\bar{3}m$ (D5) space group similar to the parent compound ($Bi_2Se_3$). The refined lattice parameters obtained are a =b= 4.146(3) Å and c = 28.670(4) Å. Clearly, the values of the refined lattice parameters obtained for Eu doped sample are close to that of the pristine one [45]. Furthermore, Fig. 2(b) shows the respective crystallographic planes (Miller indices) and confirms that the studied $Eu_{0.1}Bi_{1.9}Se_3$ sample exhibited a single phase without any impurities within the XRD limits.

To understand the surface topography and chemical stoichiometry/elemental analysis of as synthesized $Eu_{0.1}Bi_{1.9}Se_3$ sample, we collected the SEM and EDXA data, respectively as shown in Figure 3 (a-e). The SEM images of as grown $Eu_{0.1}Bi_{1.9}Se_3$ sample were taken on freshly cleaved silvery single crystal surfaces. Figure 3(a) clearly shows that the studied $Eu_{0.1}Bi_{1.9}Se_3$ sample exhibits a slab like layered structure stacked one over the other, similar to the pristine ($Bi_2Se_3$) sample. The EDXA data as displayed in Fig. 3(b) confirmed that the as grown $Eu_{0.1}Bi_{1.9}Se_3$ sample is pure (uncontaminated from impurities like carbon/oxygen) and composed of atomic constituents Eu, Bi and Se respectively. The quantitative weight% values of the atomic constituents (Eu, Bi and Se) were found to be near to stoichiometric, i.e., close to $Eu_{0.1}Bi_{1.9}Se_3$ (inset of Fig. 3b) with a small loss of selenium which may have occurred during the preparation of the sample. As mentioned in the introduction, that crystal growth is itself a very challenging task, so a perfect crystal cannot be achieved, i.e., some elemental loss do occur during synthesis process. Here, we can say that Eu is substituted at Bi site and not in the Van der Waals Layer. Also, the situation is different than as for reported $MnBi_2Te_4$ i.e. MnTe is inserted in Van der Waals Layer of $Bi_2Te_3$ [48] However, to know the actual reason it may be advisable to focus further on doping $Bi_2Se_3$ with other concentrations of Eu as well. $Bi_2Te_3$ is more susceptible for 3d magnetic layer insertion in Van der Waals Layer viz. $MnBi_2Te_4$, i.e., instead of partial doping an anti-ferromagnetic MnTe layer is inserted between $Bi_2Te_3$ unit cells. Furthermore, the elemental mapping for the $Eu_{0.1}Bi_{1.9}Se_3$ crystals [Figure 3 (c-e)] confirmed the homogeneous distribution of the constituent elements i.e., Eu, Bi and Se respectively, which is in accordance to their stoichiometric ratio.

Figure 4(a) shows the temperature dependent electrical resistivity plots at zero magnetic field (0Tesla) for pure ($Bi_2Se_3$) and Eu doped $Bi_2Se_3$ ($Eu_{0.1}Bi_{1.9}Se_3$) single crystals. ρ(T) is metallic from T=5K to about 295K for both the pristine and $Eu_{0.1}Bi_{1.9}Se_3$ sample. Moreover, pure ($Bi_2Se_3$) sample exhibits greater resistivity in comparison to the Eu doped $Bi_2Se_3$ ($Eu_{0.1}Bi_{1.9}Se_3$) sample. Inset of figure 4(a) displays the normalized conductivity (σ/ $σ_{295}$) as a function of temperature from 5K to 295K for both pristine and Eu doped $Bi_2Se_3$ samples respectively. Inset of figure 4(a) confirms the fact that though both $Bi_2Se_3$ and $Eu_{0.1}Bi_{1.9}Se_3$ samples exhibit metallic nature, the relative conductivity is less for the Eu doped $Bi_2Se_3$ sample. The resistivity as well as conductivity behaviour of pure and Eu doped $Bi_2Se_3$ crystals are in agreement to our previously reported literature [44]. Furthermore, we can say that doping $Bi_2Se_3$ with magnetic (Eu and Co) impurities results into lesser conductivity due to the breaking of TRS. Ihe Boltzmann conductivity analysis for heavily 3d metal doped topological insulators is proposed theoretically [13, 14]. We could not perform the Boltzmann conductivity analysis as our lightly doped ($Eu_{0.1}Bi_{1.9}Se_3$) crystal is yet highly metallic. Instead the HLN analysis of the pure and Eu doped $Bi_2Se_3$ is provided in next sections.

Figure 4 (b-c) depicts the temperature dependence of electrical resistivity (ρ) for freshly cleaved $Eu_{0.1}Bi_{1.9}Se_3$ and $Bi_2Se_3$ single crystals measured under varying applied magnetic fields viz., 0,1,3 and 5Tesla. The values of resistivity for both pure and Eu doped $Bi_2Se_3$ crystals under different applied magnetic fields are observed to increase with increase in temperature from 5K to 75K. Similar to the parent compound ($Bi_2Se_3$), the ρ(T)H curves for the as grown $Eu_{0.1}Bi_{1.9}Se_3$ crystal exhibits a metallic behaviour. However, the values of resistivity for the pure crystal are much more in comparison to the Eu doped $Bi_2Se_3$ crystals. To further investigate the relative impact of magnetic field on resistivity for both pure and Eu doped $Bi_2Se_3$ crystals, we show the isothermal magneto-resistance (RH) behaviour at varying temperatures (5, 10, 20, 50 and 100K) as discussed in next section.

In particular, the MR is important for technological applications and is defined as the change in the resistivity of the material with applied magnetic fields i.e., MR (%) = {[ρ(H) - ρ(0)] / ρ(0)}*100, where H is the applied perpendicular magnetic field, ρ(H) and ρ(0) are the resistivity with and without applied magnetic field respectively. Figure 5 (a-b) displays the perpendicular magnetic field induced MR% for both pure ($Bi_2Se_3$) and Eu doped $Bi_2Se_3$ ($Eu_{0.1}Bi_{1.9}Se_3$) single crystals measured at varying temperatures (5 to 100K) and fields up to ± 5Tesla. Magneto-transport measurement schematic diagram is given in our own previous work on Co added $Bi_2Se_3$ TI [44].

MR (%) is observed to decrease with increase in temperature from 5 to 100K, for both pure ($Bi_2Se_3$) and Eu doped $Bi_2Se_3$ ($Eu_{0.1}Bi_{1.9}Se_3$) single crystals [Fig. 5(a, b)]. Table 1 displays the magnetic field dependent MR values at different temperatures for pure ($Bi_2Se_3$) and Eu doped $Bi_2Se_3$ ($Eu_{0.1}Bi_{1.9}Se_3$) single crystals. Pure $Bi_2Se_3$ exhibits a MR of ∼80% at the lowest temperature i.e., 5K and only ∼30% at 100K (Table 1). On the other hand, for Eu doped $Bi_2Se_3$ ($Eu_{0.1}Bi_{1.9}Se_3$) crystal, the MR at 5K is 32%; whereas the same decreases to 22% at 100K (Table 1). We can say that the MR reduces by half in case of the $Eu_{0.1}Bi_{1.9}Se_3$ sample in comparison to the $Bi_2Se_3$, which confirms the fact that the TRS is affected by the Eu doping. Above 100K, the MR becomes negligible for both pristine and doped samples. Further, it is clear from Fig. 5(a) and (b) that though in case of $Bi_2Se_3$ the MR is all positive, in case of $Eu_{0.1}Bi_{1.9}Se_3$, the same is –ve at lower fields. Particularly, at lower applied magnetic fields of H ≤ ± 0.2Tesla, the magneto-transport data of $Eu_{0.1}Bi_{1.9}Se_3$ exhibits crossover of WL (-ve) and WAL (+ve) behaviour with the change of temperature. This suggests a qualitatively different MR mechanism in $Eu_{0.1}Bi_{1.9}Se_3$ sample at different temperature, which is absent in case of the pure $Bi_2Se_3$ sample. As reported, that due to the TRS breaking gap opened at the Dirac point of the topologically surface states, magnetically doped TIs will undergo a WAL to WL crossover [49]. M. Liu et al. showed that when a TI is magnetically doped it gets transformed into a topologically trivial dilute magnetic semiconductor (DMS) [27]. Also, it has been proposed that the incorporation of magnetic impurities results into the increased disorder in the films causing localization in the electronic states (WL), which is strongly related to the field induced magnetization [29]. However, in our case the $Eu_{0.1}Bi_{1.9}Se_3$ single crystal exhibits complex crossover behaviour at lower applied magnetic fields. Briefly, at 5K WL dominates, which changes to WAL at 10K, re enters to WL state at 20 and 50K respectively, and finally drives back to WAL state at 100K. Indeed, the competition between WL and WAL behaviour suggests that the fact that Eu doping into $Bi_2Se_3$ has affected the TRS. The possible reason behind the occurrence of such complex crossover phenomenon is still unknown. However, to know the actual underlying mechanism behind the complex crossover of the localization behaviour, we need further studies viz., the magneto - transport measurements can be performed at various Eu concentrations and temperatures.

Furthermore, we fit the MC data to the 2D WAL model, i.e., Hikami Larkin Nagaoka (HLN) which is represented as [47]:

$$\Delta\sigma(H) = \sigma(H) - \sigma(0)$$

$$= -\frac{\propto e^2}{\pi h}\left[\ln(\frac{B_\varphi}{H}) - \Psi\left(\frac{1}{2} + \frac{B_\varphi}{H}\right)\right]$$

Where, $\Delta\sigma(H)$ represents the change of magneto-conductivity, α is a coefficient signifying the overall strength of the WAL, e denotes the electronic charge, h represents the Planck's constant, $\Psi$ is the digamma function, H is the applied perpendicular magnetic field, $B_\varphi = \frac{h}{8e\pi H l_\varphi}$ is the characteristic magnetic field and $l_\varphi$ is the phase coherence length. The value of α is positive for WL and negative for WAL. Also, a large negative value can be caused by WAL in the bulk and two decoupled surface states, each contributing with α = -0.5. The experimentally fitted value of α varies widely, due to the problems arising from differentiating the bulk and surface contributions clearly. As reported, α may lie between –0.4 and –1.1, for single surface state, two surface states, or intermixing between the surface and bulk states [51, 52].

We first applied the HLN formula to the MC curves of Eu doped ($Eu_{0.1}Bi_{1.9}Se_3$) sample measured at different temperatures. The HLN fitting becomes challenging at low temperature (5K) and lower fields (below say 1 Tesla) in particular for the studied $Eu_{0.1}Bi_{1.9}Se_3$ crystal, primarily due to the competition between WAL and WL and hence extraction of α value becomes difficult. Keeping this in view we tried to fit the magneto-conductivity (MC) data in available applied field range of ± 2Tesla and 10K. This is to avoid both low temperature and low field regions to get rid of competing WL and WAL being present in case of $Eu_{0.1}Bi_{1.9}Se_3$ crystal. Figure 6 shows the MC curves for both pure ($Bi_2Se_3$) and Eu doped $Bi_2Se_3$ ($Eu_{0.1}Bi_{1.9}Se_3$) single crystals at 10K with fields up to ± 2Tesla. Both pure and doped sample, exhibits α value as ~ - 0.998 and -1 respectively, signifying WAL clearly dominating the MC data with a negligible WL component. The value of phase coherence length ($l_\varphi$), obtained for pure and doped sample is 11.61 and 14.75nm respectively. Interestingly in both cases (pure and Eu doped $Bi_2Se_3$) the α and ($l_\varphi$) values are within the range of surface states dominated conduction.

**Conclusion**

We successfully grew good quality Eu doped $Bi_2Se_3$ ($Eu_{0.1}Bi_{1.9}Se_3$) single crystals which exhibited sharp reflections along 00l alignment, indicating good crystalline quality. SEM images confirmed the surface topography and composition of as grown Eu doped $Bi_2Se_3$ ($Eu_{0.1}Bi_{1.9}Se_3$) single crystals. EDXA mapping showed a homogeneous distribution revealing the close chemical stoichiometry of the synthesized ($Eu_{0.1}Bi_{1.9}Se_3$) sample. We observed a decrease in the MR value when the pristine $Bi_2Se_3$ is doped with Eu viz., from ∼80% for pure $Bi_2Se_3$ to ∼32% for $Eu_{0.1}Bi_{1.9}Se_3$ at 5K. Though the magneto-transport studies at lower applied magnetic field (H≤ ± 0.2Tesla) at 5K revealed a complex crossover between WL and WAL, at relative higher fields of ± 2Tesla and temperature 10K, the WAL is seen to be dominating.

**Acknowledgements**

The authors from CSIR-NPL would like to thank their Director NPL, India, for his keen interest in the present work. Authors thank Poonam Rani for reading the MS. S. Patnaik thanks DST-SERB project (EMR/2016/003998) for the low temperature and high magnetic field facility at JNU, New Delhi. Rabia Sultana and Ganesh Gurjar thank CSIR, India, for research fellowship. Rabia Sultana thanks AcSIR-NPL for Ph.D. registration.

**Figure Captions**

**Figure 1:** Schematic heat treatment diagram for $Eu_{0.1}Bi_{1.9}Se_3$ single crystal.

**Figure 2:** (a) X-ray diffraction pattern of as synthesized $Eu_{0.1}Bi_{1.9}Se_3$ and (b) Rietveld fitted room temperature XRD pattern for powder $Eu_{0.1}Bi_{1.9}Se_3$ crystal.

**Figure 3:** (a) SEM images taken on freshly cleaved $Eu_{0.1}Bi_{1.9}Se_3$ single crystal (b-e) EDAX mapping of $Eu_{0.1}Bi_{1.9}Se_3$ single crystal.

**Figure 4:** (a) Temperature dependent electrical resistivity under zero applied magnetic field for $Eu_{0.1}Bi_{1.9}Se_3$ and $Bi_2Se_3$ single crystal. Inset shows the temperature dependent normalized conductivity curves for both $Eu_{0.1}Bi_{1.9}Se_3$ and $Bi_2Se_3$ single crystal. Temperature dependent electrical resistivity under different applied magnetic field for (b) $Eu_{0.1}Bi_{1.9}Se_3$ and (c) $Bi_2Se_3$ single crystal.

**Figure 5:** MR (%) as a function of perpendicular applied magnetic field (H) at different temperatures for (a) $Bi_2Se_3$ (b) $Eu_{0.1}Bi_{1.9}Se_3$ single crystal.

**Figure 6:** Magneto-conductivity (MC) analysis for $Bi_2Se_3$ and $Eu_{0.1}Bi_{1.9}Se_3$ single crystal fitted using the HLN equation at 10K up to ± 2Tesla.

Table 1

| Temperature | MR (%) | |
|---|---|---|
| | $Bi_2Se_3$ (up to ± 5Tesla) | $Eu_{0.1}Bi_{1.9}Se_3$ (up to ± 5Tesla) |
| 5K | ~80 | ~32 |
| 10K | ~80 | ~38 |
| 20K | ~80 | ~35 |
| 50K | ~60 | ~30 |
| 100K | ~30 | ~22 |

Fig. 1

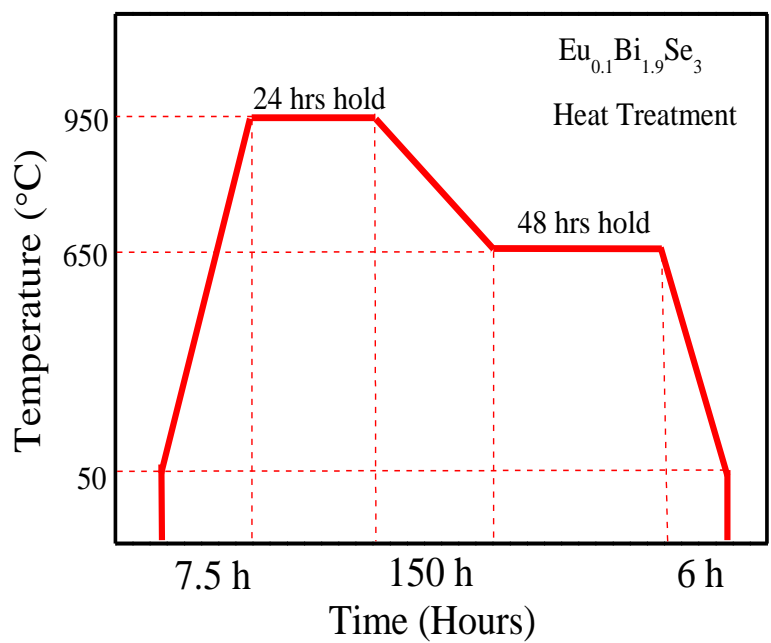

Fig. 2(a)

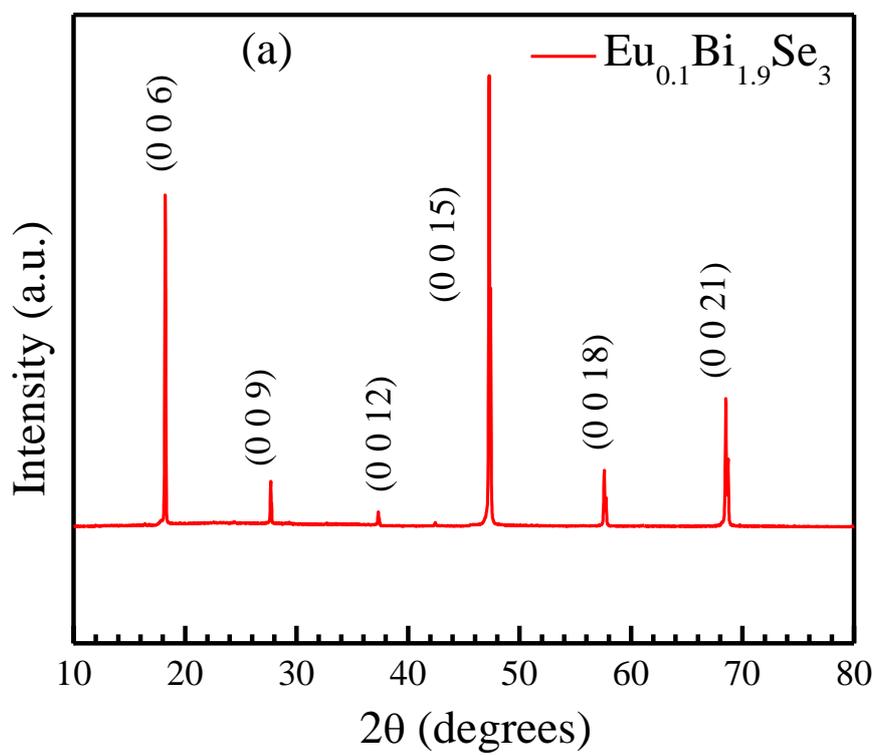

Fig. 2(b)

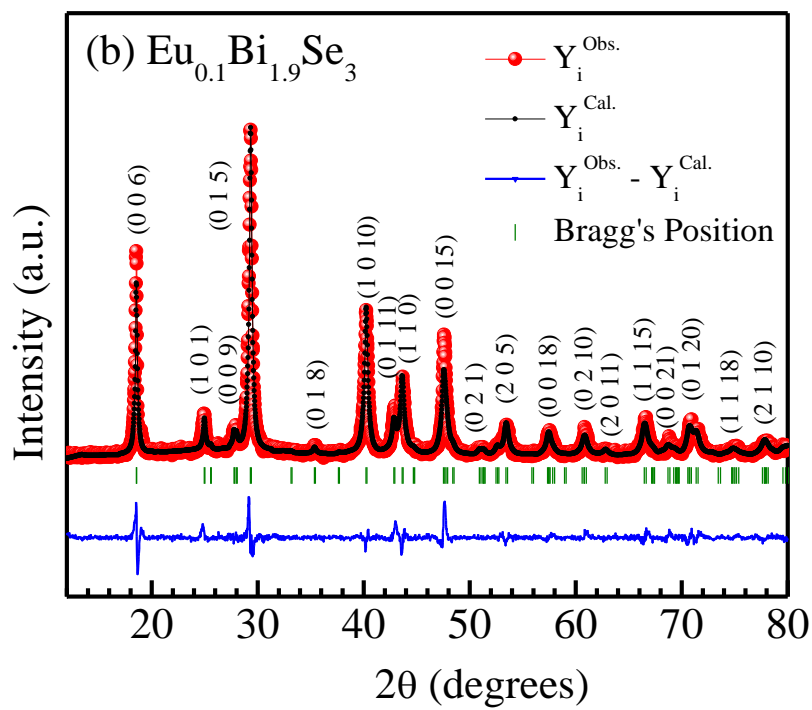

Fig. 3(a)

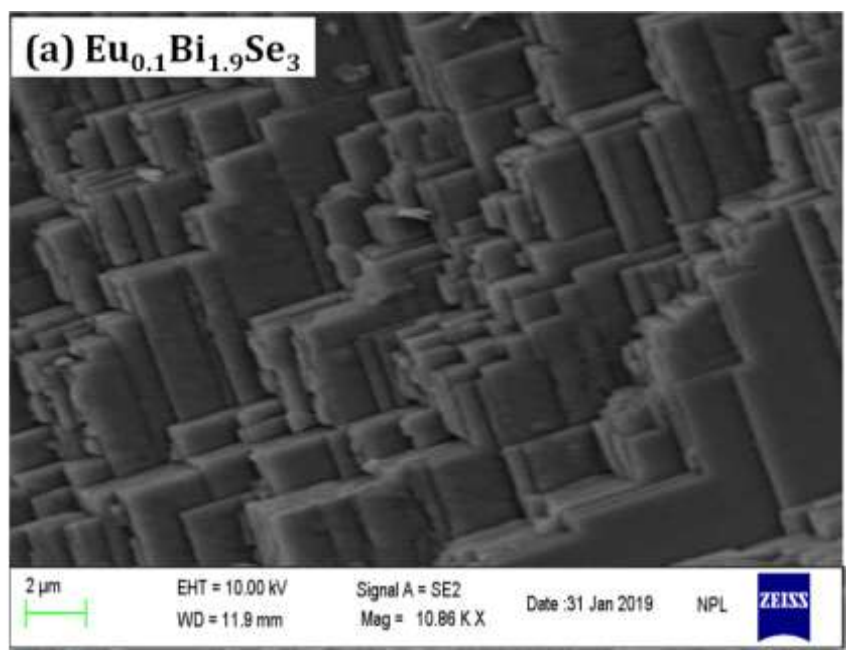

Fig. 3(b)

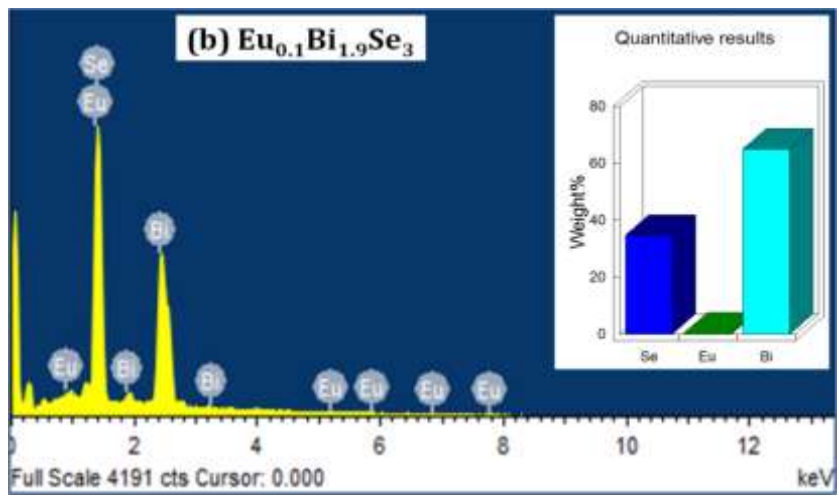

Fig. 3(c)

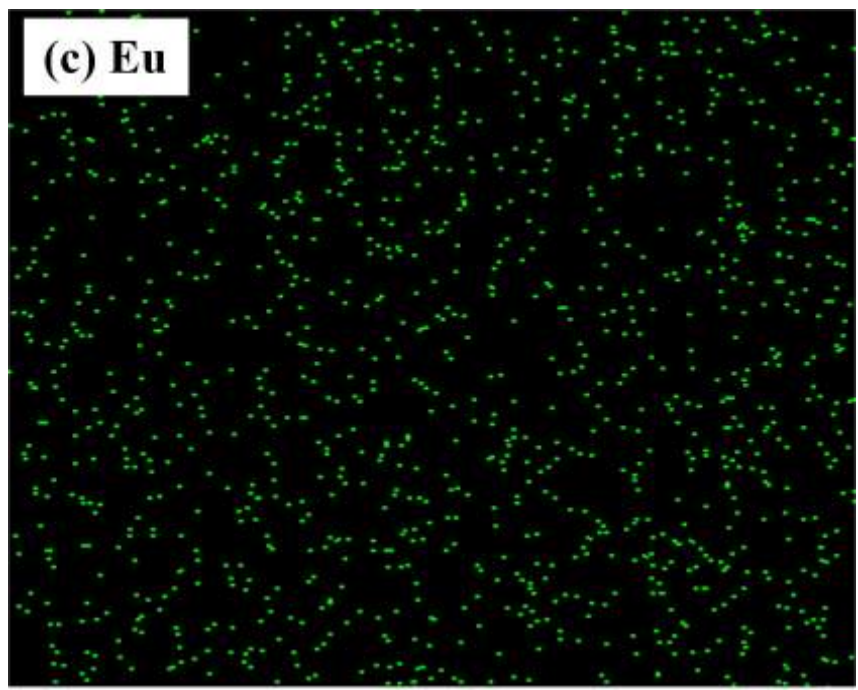

Fig. 3(d)

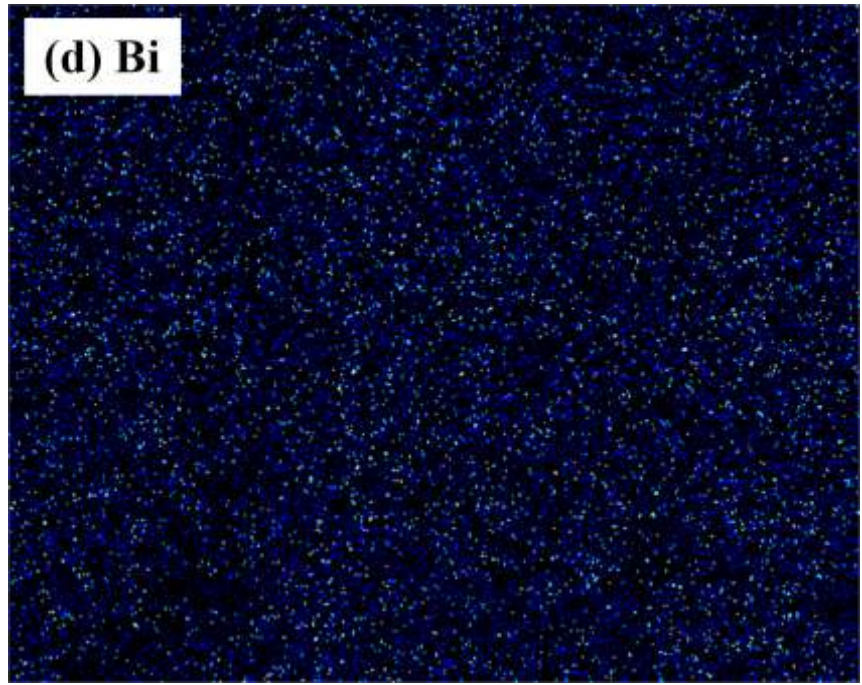

Fig. 3 (e)

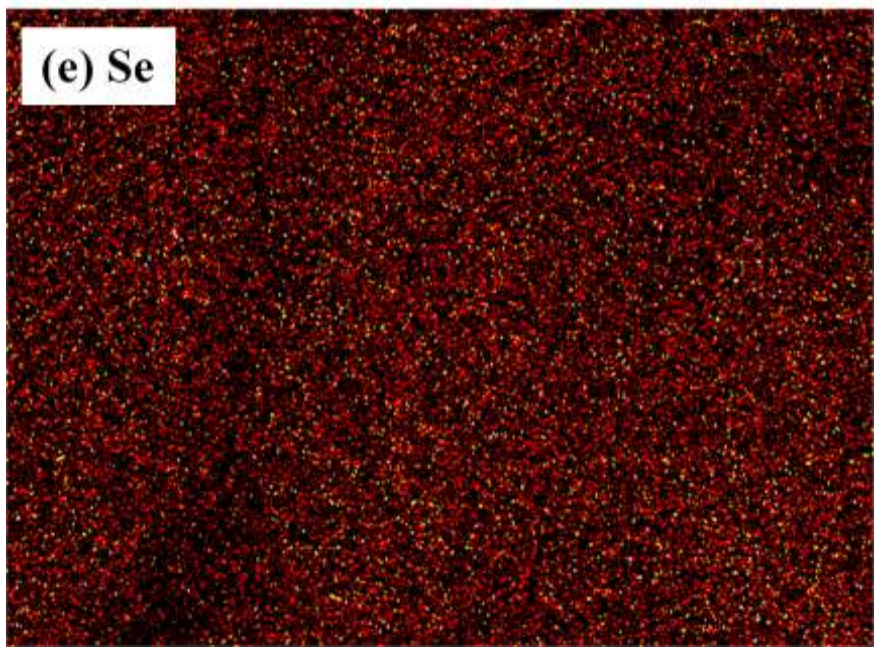

Fig. 4(a)

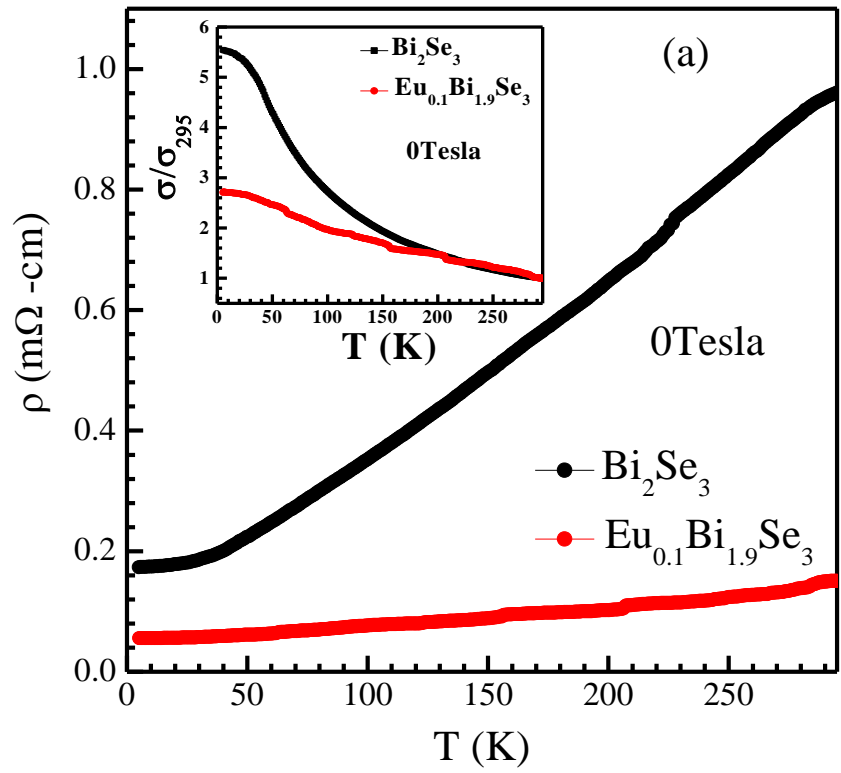

Fig. 4(b)

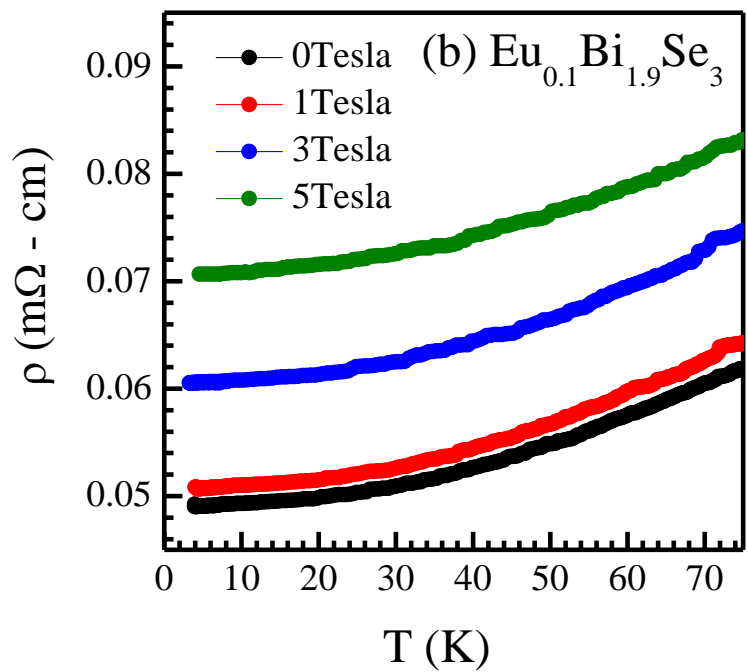

Fig. 4(c)

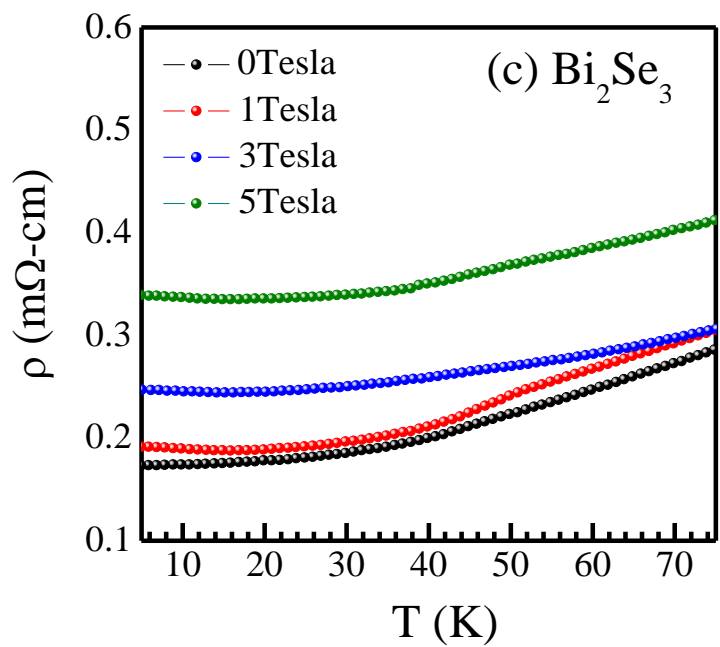

Fig. 5(a)

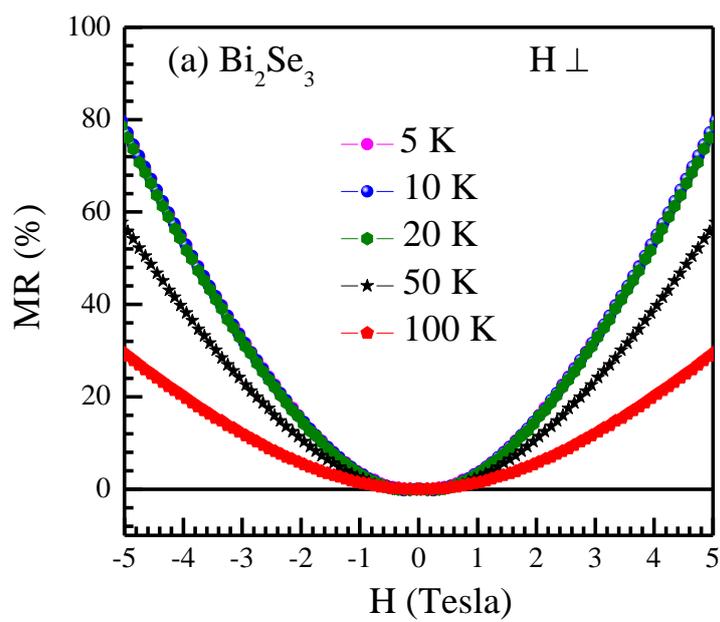

Fig. 5(b)

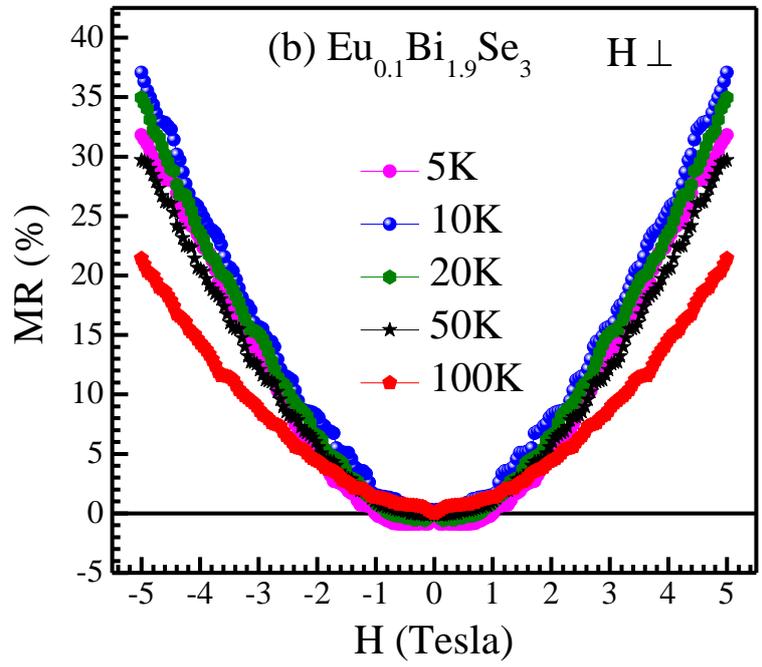

Fig. 6

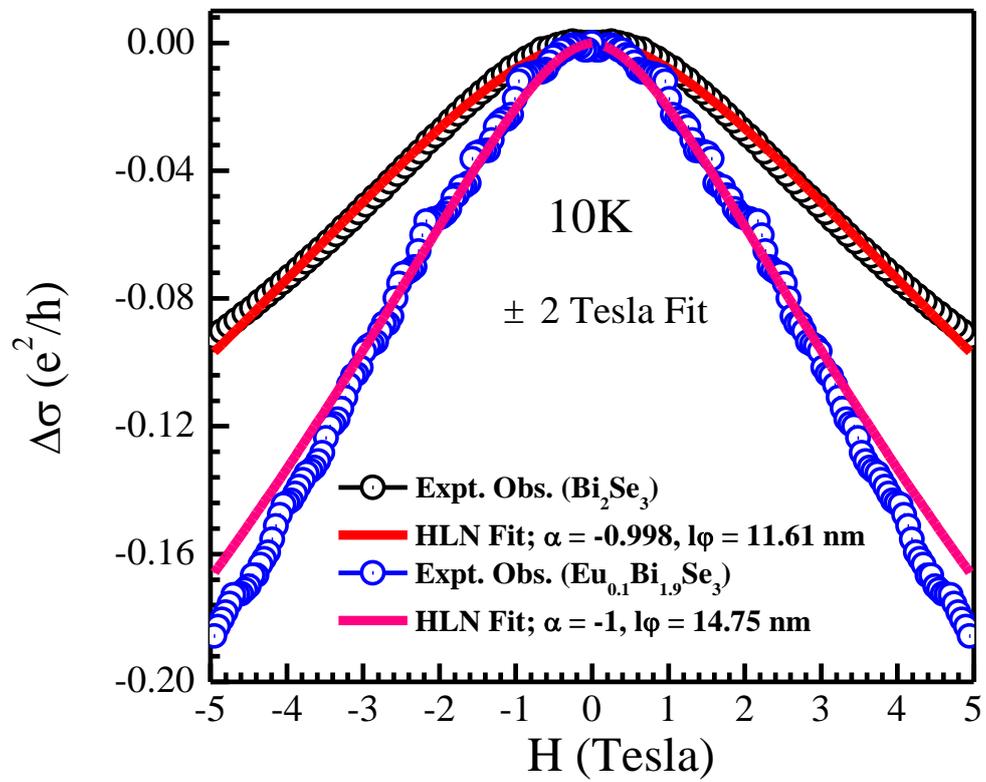